\documentclass[aps,10pt,pra,twocolumn,showpacs,nofootinbib,superscriptaddress]{revtex4-1}
\usepackage{amsfonts}
\usepackage{amsmath}
\usepackage{amssymb}
\usepackage{graphicx}%
\usepackage{color}
\usepackage{cancel}
\usepackage{ulem}
\usepackage{comment}

\def\filetype{eps}
\providecommand{\U}[1]{\protect\rule{.1in}{.1in}}
\providecommand{\U}[1]{\protect\rule{.1in}{.1in}}

\begin{document}
\preprint{ }
\title{Nonequilibrium States of a Quenched Bose Gas}
\author{Ben Kain}
\affiliation{Department of Physics, College of the Holy Cross, Worcester, Massachussets 01610, USA}
\affiliation{Department of Physics and Astronomy, Rowan University, Glassboro, New
Jersey 08028, USA}
\author{Hong Y.\ Ling}
\affiliation{Department of Physics and Astronomy, Rowan University, Glassboro, New
Jersey 08028, USA}
\affiliation{Kavli Institute for Theoretical Physics, University of California, Santa
Barbara, California 93106, USA }
\affiliation{ITAMP, Harvard-Smithsonian Center for Astrophysics, Cambridge, Massachussets 02138, USA}

\begin{abstract}
\noindent Yin and Radzihovsky \cite{yin13arXiv:1308.6376} recently developed a
self-consistent extension of a Bogoliubov theory, in which the condensate
number density $n_{c}$ is treated as a mean field that changes with time, in
order to analyze a JILA experiment by Makotyn \textit{et al.}
\cite{makotyn13arXiv:1308.3696} on a $^{85}$Rb Bose gas following a deep
quench to a large scattering length. We apply this theory to construct a
closed set of equations that highlight the role of $\dot{n}_{c}$, which is to
induce an effective interaction between quasiparticles. We show analytically
that such a system supports a steady state characterized by a constant
condensate density and a steady but periodically changing momentum
distribution, whose time average is described exactly by the generalized Gibbs
ensemble. We discuss how the $\dot{n}_{c}$-induced effective interaction,
which cannot be ignored on the grounds of the adiabatic approximation for
modes near the gapless Goldstone mode, can significantly affect condensate
populations and Tan's contact for a Bose gas that has undergone a deep quench.

\end{abstract}


\maketitle

\section{Introduction}

The recent explosive interest in using ultracold atomic gases as an excellent
platform for studying the dynamics of strongly correlated systems driven
out-of-equilibrium by slow (adiabatic) or sudden (quenched) changes to system
parameters
\cite{bloch08RevModPhys.80.885,polkovnikov11RevModPhys.83.863,cazalilla11RevModPhys.83.1405}
has been fueled by the unprecedented ability to tune such system parameters,
in particular the interatomic interaction \cite{chin10RevModPhys.82.1225}, and
early experimental
\cite{greiner02Nature.415.39,donley02Nature.417.529,sadler06Nature.443.312,kinoshita06Nature.440.900}
and theoretical
\cite{kokkelmans02PhysRevLett.89.180401,barankov04PhysRevLett.93.160401,andreev04PhysRevLett.93.130402,yuzbashyan05PhysRevB.72.220503,yuzbashyan06PhysRevLett.96.097005,rigol07PhysRevLett.98.050405,cazalilla06PhysRevLett.97.156403}
explorations of nonequilibrium dynamics.  At the forefront of such studies
are questions regarding nonequilibrium states reached after quenching
\cite{polkovnikov11RevModPhys.83.863}, whose observation hinges on the ability
of many-body systems to maintain coherence on time scales much longer than the
equilibration time of the particles in the systems.  The main obstacle
preventing cold toms, especially those with a large scattering length $a$,
from acquiring such a long coherence time is three-body recombination, in
which two atoms in a trap form a diatomic molecule with a third atom escaping
from the trap, causing losses.  This poses less of a challenge in fermionic
systems, thanks to the Pauli exclusion principle which tends to suppress
three-body recombination \cite{petrov04PhysRevLett.93.090404}.  As such, in
Fermi gases tuned across or close to the Feshbach resonance, where $a$ goes to
infinity, researchers have observed, among other things, the crossover from
Fermi to Bose superfluids
\cite{bartenstein04PhysRevLett.92.120401,regal04PhysRevLett.92.040403,zwierlein05Nature.435.1047}%
, rich phase separation scenarios \cite{partridge06Science.311.503}, and
universalities of Fermi gases
\cite{kinast05Science.307.1296,stewart06PhysRevLett.97.220406,horikoshi10Science.327.442,nascimbene10Nature.463.1057,ku12Science.335.563}%
.

By contrast, we are less fortunate with bosonic systems. In the weak
interaction limit, the three-body loss rate $\propto n^{2}a^{4}$
\cite{fddichev96PhysRevLett.77.2921} increases with $a$ $\left(  >0\right)
$ far faster than the equilibration rate $\propto na^{2}v$, where $n$ is the
atom number density and $v$ is the average velocity. In the strong
interaction limit, systems are highly nonlinear.  In the extreme case of
unitarity, where $a\rightarrow\infty$, the interatomic distance, $n^{-1/3}$,
remains the only physically relevant length scale,\ and the unitary Bose gas
\cite{cowell02PhysRevLett.88.210403,lee10PhysRevA.81.063613,diederix11PhysRevA.84.033618,li12PhysRevLett.108.195301,borzov12PhysRevA.85.023620,jiang13arXiv:1307.4263,zhou13AnnalsPhys.328.83,stoof13arXiv:1302.1792,stoof14JLowTemp174.159}
is expected to display universal properties akin to those of a Fermi gas at
unitarity
\cite{ho04PhysRevLett.92.090402,nikolic07PhysRevA.75.033608,veillette07PhysRevA.75.043614}%
. The prospect of a well-defined unitary limit was brightened by experiments
on dilute thermal Bose gases
\cite{rem13PhysRevLett.110.163202,fletcher13PhysRevLett.111.125303}. At
unitarity, on purely dimensional grounds, both the three-body loss and the
equilibration rates will be of the same order of magnitude as the Fermi energy
\cite{ho04PhysRevLett.92.090402}, $\epsilon_{F}=\hbar(  \omega_{F}=\hbar
k_{F}^{2}/2m)  $, where $k_{F}=(  6\pi^{2}n)  ^{1/3}$ is the
Fermi momentum.\ It was unclear which dominates until recently when a
JILA\ experiment on $^{85}$Rb by Makotyn \textit{et al.}
\cite{makotyn13arXiv:1308.3696} firmly established that the three-body loss
rate is much slower than the equilibration rate. This pleasant surprise
opens the door to the possibility of exploring the rich physics underlying
strongly interacting Bose gases
\cite{cowell02PhysRevLett.88.210403,lee10PhysRevA.81.063613,diederix11PhysRevA.84.033618,li12PhysRevLett.108.195301,borzov12PhysRevA.85.023620,jiang13arXiv:1307.4263,zhou13AnnalsPhys.328.83,stoof13arXiv:1302.1792,stoof14JLowTemp174.159}
and has motivated, together with experimental works such as
\cite{hung13Science.341.1213}, a flurry of theoretical studies concerning
quenched nonequilibrium dynamics
\cite{yin13arXiv:1308.6376,sykes13arXiv:1309.0828,rancon13PhysRevA.88.031601,laurent13arXiv:1312.0079,smith14PhysRevLett.112.110402,rancon14PhysRevA.90.021602,corson14arXiv:1409.0524}.

Of particular relevance here is a theoretical analysis of the JILA\ experiment
by Yin and Radzihovsky \cite{yin13arXiv:1308.6376} who, in the spirit of
Bogoliubov mean-field theory, divided the Bose gas at zero temperature into a
quantum system of Bogoliubov quasiparticles and a condensate of number density
$n_{c}$, which is treated as a dynamical mean field (as opposed to a fixed
constant as done by Natu and Mueller \cite{natu13PhysRevA.87.053607}). We
explore this same topic using this same approach. In order to distinguish as
well as highlight our work, we point out an important feature inherent to any
Bogoliubov inspired mean-field theories: mean fields ($n_{c}$ here), which act
like control (albeit self-generated) parameters to the quantum system
describing quasiparticles, change with time even \textit{after} quenching.
\ This may be contrasted with many existing models, particularly those in one
dimension (1D) to which powerful techniques such as bosonization are
accessible \cite{cazalilla11RevModPhys.83.1405}, where the control parameters,
after quenching, are all fixed independent of time. 

Our paper is organized as follows.  In Sec.\ II, we describe our model and derive the set of closed equations
(\ref{dot c_k d_k}) and (\ref{dot nc}), which depend explicitly not only on
$n_{c}$ but also on $\dot{n}_{c}$, thereby allowing us to identify that the
role of $\dot{n}_{c}$ is to induce between quasiparticles an (imaginary)
effective interaction proportional to $\dot{n}_{c}$. 

In Sec.\ III, we apply the theory developed in Sec.\ II to address several
questions which are of both theoretical and experimental interest. Will such
a system reach a steady state? The answer seems affirmative from both the
measurement in experiment \cite{makotyn13arXiv:1308.3696} and the numerical
investigation in theory \cite{yin13arXiv:1308.6376}. Here, we show
\textit{analytically} that this is true in the thermodynamic limit.  What
are the properties of such a steady state? This stationary state is
characterized by a time-independent condensate density and a steady but
periodically changing momentum distribution of quasiparticles. The authors of
the recent JILA\ experiment \cite{makotyn13arXiv:1308.3696}, while recognizing
the excellent fit between measured momentum and an ideal Bose gas
distribution, were nevertheless open to other possibilities. We show
\textit{analytically} that the time average of this distribution is described
exactly by the generalized Gibbs ensemble. 

In Sec.\ IV, we study the adiabatic solution obtained in the absence of the
$\dot{n}_{c}$-induced interaction. We stress and demonstrate that our
formalism, where the role of $\dot{n}_{c}$ is explicitly built in, lends
itself naturally to the adiabatic theorem, allowing us to estimate
straightforwardly the effect of the $\dot{n}_{c}$-induced interaction on the
system dynamics.

In Sec.\ V, we perform a detailed study of the condensate population dynamics
at both short and long times. We apply a self-consistent perturbation
theory, which we develop in the Appendix, to quantify the short-time behavior
of the adiabatic solution. We use the adiabatic theorem to explain why the
$\dot{n}_{c}$-induced interaction can significantly slow down the initial
condensate dynamics so that $n_{c}(  t)  $ approaches a finite
number even when the Bose gas is quenched to unitarity. 

In Sec.\ VI, we quantify the \textquotedblleft contact\textquotedblright\ at
both short and long times\ in connection with the quasiparticle distribution
at high momenta, where the \textquotedblleft contact\textquotedblright\ is the
central theme that unites several universal relations that Tan discovered from a study of
Fermi gases with zero-range interaction
\cite{tan08AnnPhys.323.2952,tan08AnnPhys.323.2971,tan08AnnPhys.323.2987}.

We summarize our results in Sec.\ VII.

\section{Models and Equations}

The study of a strongly interacting Bose gas, where the underlying
physics involves an intricate interplay between few-body systems and many-body
backgrounds, is a complex endeavor that goes beyond the scope of the present
work if no restrictions are imposed on the model.  In particular, three-body
collisions can lead to the formation of dimers and (Efimov) trimers
\cite{braaten06PhysRept.428.259} so that the state of atomic BEC is at best
metastable. In this work, we ignore the three-body related effects and focus
on the physics of this metastable sate. This practice is now supported by
the results of JILA's experiment, which demonstrated that quenching to a large
scattering length can create metastable BECs that exist for a sufficiently
long time. 

Following Ref.\ \cite{yin13arXiv:1308.6376}, we model the interacting
Bose gas (with a broad Feshbach resonance) by a single-channel
grand-canonical Hamiltonian,
\begin{equation}
\hat{H}=\int d^{3}\mathbf{r}\, \hat{\psi}^{\dag}\left(  \mathbf{r}\right)
\hat{h}_{0}\hat{\psi}\left(  \mathbf{r}\right)  +\frac{g}{2}\int
d^{3}\mathbf{r}\, \hat{\psi}^{\dag2}\left(  \mathbf{r}\right)  \hat{\psi}%
^{2}\left(  \mathbf{r}\right)  , \label{Hamiltonian in position space}%
\end{equation}
where $\hat{\psi}\left(  \mathbf{r}\right)  $ is the bosonic field operator in
position\ space, $\hat{h}_{0}=-\hbar^{2}\nabla^{2}/2m-\mu$ is the
single-particle Hamiltonian operator with $\mu$ the chemical potential, and
$g=4\pi\hbar^{2}a/m$ measures the two-body contact interaction with $a$ the
$s$-wave scattering length.

The dynamics of a system described by the
Hamiltonian in Eq.\ (\ref{Hamiltonian in position space}) following a sudden
quench where $a$ ($g)$ changes abruptly from $a_{i}$ $(  g_{i})  $
to $a_{f}$ $(  g_{f})  $ can, in general, be quite complicated. As a simplification, we assume that the system remains, throughout its
evolution, in a state where the Bogoliubov perturbative ansatz, $\hat{\psi
}\left(  \mathbf{r}\right)  =\psi+\delta\hat{\psi}\left(  \mathbf{r}\right)
$, remains valid. Here, as in equilibrium problems, $\delta\hat{\psi}\left(
\mathbf{r}\right)  =\sum_{\mathbf{k}\neq0}\hat{a}_{\mathbf{k}}e^{i\mathbf{k}%
\cdot\mathbf{r}}/\sqrt{V}$ with $V$ the total volume, is an operator
describing fluctuations around the macroscopic (but uniform) condensate of
density $n_{c}=| \psi| ^{2}$ and chemical potential
$\mu=gn_{c}$, where $n_{c}$ is treated as a dynamical mean field,
$n_{c}\left(  t\right)  $.

\ The Hamiltonian after quenching $\left(  t>0^{+}\right)  $, up to second
order in $\hat{a}_{\mathbf{k}}$, then reads $\hat{H}=-Vg_{f}n_{c}^{2}%
/2+\hat{H}_{2}$, where
\begin{equation}
\hat{H}_{2}=\sum_{\mathbf{k}\neq0}\left[  \left(  \epsilon_{k}+g_{f}%
n_{c}\right)  \hat{a}_{\mathbf{k}}^{\dag}\hat{a}_{\mathbf{k}}+\frac{g_{f}%
n_{c}}{2}\left(  \hat{a}_{\mathbf{k}}^{\dag}\hat{a}_{-\mathbf{k}}^{\dag}%
+\hat{a}_{\mathbf{k}}\hat{a}_{-\mathbf{k}}\right)  \right]  , \label{H2}%
\end{equation}
with $\epsilon_{k}=\hbar^{2}k^{2}/2m$ the kinetic energy of an atom. Note
that the same $\hat{H}$, with $g_{f}$ replaced with $g_{i}$, describes a
system in equilibrium before quenching $\left(  t<0^{-}\right)  $. A
quadratic Hamiltonian like Eq.\ (\ref{H2}) can be diagonalized into
\[
\hat{H}=H_{0}+\sum_{\mathbf{k}\neq0}E_{k}\left(  t\right)  \hat{\gamma
}_{\mathbf{k}}^{\dag}\left(  t\right)  \hat{\gamma}_{\mathbf{k}}\left(
t\right)  ,
\]
[with $H_{0}=-V\frac{g_{f}n_{c}^{2}}{2}+\frac{1}{2}\sum_{\mathbf{k}\neq
0}(  E_{k}-\epsilon_{k}-g_{f}n_{c})  $] by means of a Bogoliubov
transformation,
\begin{equation}
\left[
\begin{array}
[c]{c}%
\hat{a}_{\mathbf{k}}\\
\hat{a}_{-\mathbf{k}}^{\dag}%
\end{array}
\right]  =\left[
\begin{array}
[c]{cc}%
u_{k} & -v_{k}\\
-v_{k} & u_{k}%
\end{array}
\right]  \left[
\begin{array}
[c]{c}%
\hat{\gamma}_{\mathbf{k}}\\
\hat{\gamma}_{-\mathbf{k}}^{\dag}%
\end{array}
\right]  , \label{Bogoliubov Transformation}%
\end{equation}
where
\begin{equation}%
\begin{array}
[c]{c}%
u_{k}\left(  t\right) \\
v_{k}\left(  t\right)
\end{array}
=\sqrt{\frac{1}{2}\left(  \frac{\epsilon_{k}+g_{f}n_{c}\left(  t\right)
}{E_{k}\left(  t\right)  }\pm1\text{ }\right)  } \label{uv}%
\end{equation}
are the Bogoliubov parameters and
\begin{equation}
E_{k}\left(  t\right)  =\sqrt{\epsilon_{k}\left(  \epsilon_{k}+2g_{f}%
n_{c}\left(  t\right)  \right)  } \label{E_k}%
\end{equation}
is the quasi-particle energy dispersion.

The time evolution of the system is governed by the Heisenberg equation of
motion for operator $\hat{\gamma}_{\mathbf{k}}$,
\begin{equation}
\frac{d\hat{\gamma}_{\mathbf{k}}}{dt}=-\frac{i}{\hbar}\left[  \hat{\gamma
}_{\mathbf{k}},\hat{H}_{2}\right]  +\frac{\partial\hat{\gamma}_{\mathbf{k}}%
}{\partial t}. \label{Heisenberg's equation}%
\end{equation}
A comment is in order. The original field operator $\hat{a}_{\mathbf{k}}$ in
Hamiltonian (\ref{H2}), by definition, does not depend on time explicitly.
\ Then, the quasiparticle field operator $\hat{\gamma}_{\mathbf{k}}$
introduced through the Bogoliubov transformation
(\ref{Bogoliubov Transformation}) depends explicitly on time via $n_{c}(
t)  $, which is treated as a dynamical mean field in the generalized
Bogoliubov theory. As a result, the second term $\partial\hat{\gamma
}_{\mathbf{k}}/\partial t$ should be present in the Heisenberg equation of
motion. 

Applying the Bogoliubov transformation in Eq.\ (\ref{Bogoliubov Transformation}%
) along with the time derivatives of Eqs.\ (\ref{uv}), we change Eq.\
(\ref{Heisenberg's equation}) into%
\begin{equation}
\frac{d\hat{\gamma}_{\mathbf{k}}}{dt}=-\frac{i}{\hbar}E_{k}\left(  t\right)
\hat{\gamma}_{\mathbf{k}}+g_{f}\dot{n}_{c}\left(  t\right)  \frac{\epsilon
_{k}}{2E_{k}^{2}\left(  t\right)  }\hat{\gamma}_{-\mathbf{k}}^{\dag},
\label{Heisenberg's Equation 1}%
\end{equation}
where the final term would have been absent had we ignored the partial
derivative in Eq.\ (\ref{Heisenberg's equation}). Note that Eq.\
(\ref{Heisenberg's Equation 1}) would be the Heisenberg equation of the
Hamiltonian $\hat{H}^{\prime}=\sum_{\mathbf{k}\neq0}[  E_{k}(
t)  \hat{\gamma}_{\mathbf{k}}^{\dag}\hat{\gamma}_{\mathbf{k}}%
+U_{k}(  t)  (  \hat{\gamma}_{\mathbf{k}}^{\dag}\hat{\gamma
}_{-\mathbf{k}}^{\dag}+\hat{\gamma}_{\mathbf{k}}\hat{\gamma}_{-\mathbf{k}%
})  ]  $ if $\hat{\gamma}_{\mathbf{k}}$ were treated as an
operator that does not depend on time explicitly. Thus, we see that to some
extent, $\dot{n}_{c}(  t)  \neq0$ is to induce between
quasiparticles an imaginary effective interaction%
\begin{equation}
U_{k}\left(  t\right)  =i\hbar g_{f}\dot{n}_{c}\left(  t\right)  \epsilon
_{k}/2E_{k}^{2}\left(  t\right)  . \label{Ut)}%
\end{equation}

We now change the dynamical variables from operators, $\hat{\gamma
}_{\mathbf{k}}(  t)  $ and $\hat{\gamma}_{-\mathbf{k}}^{\dag
}(  t)  $, to complex numbers, $c_{k}(  t)  $ and
$d_{k}(  t)  $, via the transformation%
\begin{equation}
\left[
\begin{array}
[c]{c}%
\hat{\gamma}_{\mathbf{k}}\left(  t\right) \\
\hat{\gamma}_{-\mathbf{k}}^{\dag}\left(  t\right)
\end{array}
\right]  =\left[
\begin{array}
[c]{cc}%
c_{k}\left(  t\right)  & -d_{k}^{\ast}\left(  t\right) \\
-d_{k}\left(  t\right)  & c_{k}^{\ast}\left(  t\right)
\end{array}
\right]  \left[
\begin{array}
[c]{c}%
\hat{\alpha}_{\mathbf{k}}\\
\hat{\alpha}_{-\mathbf{k}}^{\dag}%
\end{array}
\right]  , \label{c_k and d_k}%
\end{equation}
where $| c_{k}(  t)  | ^{2}-|
d_{k}(  t)  | ^{2}=1$ and $\hat{\alpha}_{\mathbf{k}}$ is
the quasi-particle operator defined with respect to the pre-quench vacuum
$\vert 0^{-}\rangle $. $\hat{\gamma}_{\mathbf{k}}(  t)
$ thus defined is a solution to Eq.\ (\ref{Heisenberg's Equation 1}) provided
that%
\begin{equation}
\frac{d}{dt}\left[
\begin{array}
[c]{c}%
c_{k}\\
d_{k}%
\end{array}
\right]  =\left[
\begin{array}
[c]{cc}%
-\frac{i}{\hbar}E_{k}\left(  t\right)  & -\frac{\epsilon_{k}g_{f}}{2E_{k}%
^{2}\left(  t\right)  }\dot{n}_{c}\left(  t\right) \\
-\frac{\epsilon_{k}g_{f}}{2E_{k}^{2}\left(  t\right)  }\dot{n}_{c}\left(
t\right)  & +\frac{i}{\hbar}E_{k}\left(  t\right)
\end{array}
\right]  \left[
\begin{array}
[c]{c}%
c_{k}\\
d_{k}%
\end{array}
\right]  , \label{dot c_k d_k}%
\end{equation}
where $c_{k}$ and $d_{k}$ are subject to the initial condition
\begin{equation}
c_{k}\left(  0^{+}\right)  =u_{k}^{+}u_{k}^{-}-v_{k}^{+}v_{k}^{-},
\quad
d_{k}\left(  0^{+}\right)  =u_{k}^{+}v_{k}^{-}-v_{k}^{+}u_{k}^{-}.
\label{initial condition 1}%
\end{equation}

In a quench experiment where the quench parameter is changed from an initial
to a final value very rapidly, specifically in a time much shorter than any
other characteristic time scale of the system, one can apply the so-called
\textit{sudden} approximation in which the state of the system immediately
after quenching at $t=0^{+}$ is assumed to be same as that immediately before
quenching, namely $\hat{a}_{\mathbf{k}}(  0^{+})  =\hat
{a}_{\mathbf{k}}(  0^{-})  $ and $n_{c}(  0^{+})
=n_{c}(  0^{-})  \equiv n_{i}$. The initial condition in Eq.\
(\ref{initial condition 1}) is derived from this sudden approximation, where
$u_{k}^{-},v_{k}^{-}$, and $E_{k}^{-}$ ($u_{k}^{+},v_{k}^{+}$, and $E_{k}^{+}%
$) are the same as Eqs.\ (\ref{uv}) and (\ref{E_k}) except that $g_{f}$ and
$n_{c}\left(  t\right)  $ are set to their $t=0^{-}$ ($t=0^{+}$) values or
explicitly,
\begin{equation}%
\begin{array}
[c]{c}%
u_{k}^{-}\\
v_{k}^{-}%
\end{array}
=\sqrt{\frac{1}{2}\left(  \frac{\epsilon_{k}+g_{i}n_{i}}{E_{k}^{-}}\text{ }%
\pm1\right)  },
\end{equation}
where
\begin{equation}
E_{k}^{-}=\sqrt{\epsilon_{k}\left(  \epsilon_{k}+2g_{i}n_{i}\right)  },
\end{equation}
and%
\begin{equation}%
\begin{array}
[c]{c}%
u_{k}^{+}\\
v_{k}^{+}%
\end{array}
=\sqrt{\frac{1}{2}\left(  \frac{\epsilon_{k}+g_{f}n_{i}}{E_{k}^{+}}\pm1\text{
}\right)  },
\end{equation}
where
\begin{equation}
E_{k}^{+}=\sqrt{\epsilon_{k}\left(  \epsilon_{k}+2g_{f}n_{i}\right)  }.
\end{equation}

The condensate density $n_{c}(  t)  $ in Eq.\ (\ref{dot c_k d_k}) is
to be obtained self-consistently from total particle number conservation,
\begin{equation}
n_{c}\left(  t\right)  =n-\left[  n_{d}\left(  t\right)  \equiv V^{-1}%
\sum_{\mathbf{k}}n_{k}\left(  t\right)  \right]  , \label{number equation}%
\end{equation}
where $n_{d}(  t)  $ is the quasiparticle population or the
condensate depletion and $n_{k}(  t)  =\langle \hat
{a}_{\mathbf{k}}^{\dag}(  t)  \hat{a}_{\mathbf{k}}(  t)
\rangle $ or
\begin{equation}
n_{k}\left(  t\right)  =\left\vert u_{k}\left(  t\right)  d_{k}\left(
t\right)  +v_{k}\left(  t\right)  c_{k}\left(  t\right)  \right\vert ^{2},
\label{nk(t) 1}%
\end{equation}
is the momentum distribution of quasiparticles (or noncondensed particles).
(Throughout, averages like $\langle \hat{A}\rangle $ will always be
defined with respect to the pre-quench vacuum: $\langle 0^{-}|
\hat{A}| 0^{-}\rangle $.) From the time derivative of the
number equation (\ref{number equation}), it follows that
\begin{equation}
\frac{dn_{c}\left(  t\right)  }{dt}=-g_{f}n_{c}\frac{1}{V}\sum_{\mathbf{k}%
\neq0}\left\{  D_{k}-\frac{2}{\hbar}\operatorname{Im}\left[  c_{k}^{\ast}%
d_{k}\right]  \right\}  , \label{dnc(t)}%
\end{equation}
where
\begin{align}
D_{k}    &= 2\left(  \dot{v}_{k}-\frac{\epsilon_{k}g_{f}\dot{n}_{c}u_{k}%
}{2E_{k}^{2}}\right)  \left[  v_{k}\left\vert c_{k}\right\vert ^{2}%
+u_{k}\operatorname{Re}\left(  c_{k}^{\ast}d_{k}\right)  \right]  \nonumber\\
&\quad + 2\left(  \dot{u}_{k}-\frac{\epsilon_{k}g_{f}\dot{n}_{c}v_{k}}{2E_{k}^{2}%
}\right)  \left[  u_{k}\left\vert d_{k}\right\vert ^{2}+v_{k}\operatorname{Re}%
\left(  d_{k}^{\ast}c_{k}\right)  \right]  .
\end{align}
Finally, with the help of the time derivatives of Eqs.\ (\ref{uv}), we arrive
at%
\begin{equation}
\frac{dn_{c}\left(  t\right)  }{dt}=2\frac{g_{f}n_{c}}{\hbar}\frac{1}{V}%
\sum_{\mathbf{k}\neq0}\operatorname{Im}\left[  c_{k}^{\ast}\left(  t\right)
d_{k}\left(  t\right)  \right]  , \label{dot nc}%
\end{equation}
which, together with Eq.\ (\ref{dot c_k d_k}), forms a closed set of equations.
\ In what follows, we use Eqs.\ (\ref{dot c_k d_k}) and (\ref{dot nc}) to study
quenched dynamics. 

\section{Nonequilibrium State and Its Properties\ }

We shall first show self-consistently that in the limit $t\rightarrow\infty$ Eqs.\ (\ref{dot c_k d_k}) and (\ref{dot nc}) support a steady-state characterized by $\dot{n}_{c}(
t)  \approx0$ or $n_{c}(  t)  \approx n_{c}^{s}$, where a
variable with superscript $s$ denotes the steady state value. The proof will
make an explicit use of $dn_{c}(  t)  /dt$ in  Eq.\ (\ref{dot nc}), illustrating that the theory we
developed in the previous section is indispensable to the nonequilibrium
problem under consideration. The proof begins with the assumption that
$\dot{n}_{c}\approx0$, which then leads, from Eq.\ (\ref{dot c_k d_k}), to
\begin{equation}
c_{k}\left(  t\right)  \approx c_{k}^{s}\exp\left(  -\frac{i}{\hbar}E_{k}%
^{s}t\right)  ,
\quad
d_{k}\left(  t\right)  \approx d_{k}^{s}\exp\left(
\frac{i}{\hbar}E_{k}^{s}t\right)  ,
\end{equation}
where $E_{k}^{s}$ is Eq.\ (\ref{E_k}) when $n_{c}$ is replaced with $n_{c}^{s}%
$. To check the self-consistency, we substitute these long time $c_{k}(
t)  $ and $d_{k}(  t)  $ into Eq.\ (\ref{dot nc}), which
results in
\begin{equation}
\dot{n}_{c}\left(  t\right)  \propto\frac{1}{V}\sum_{\mathbf{k}\neq
0}\operatorname{Im}\left[  c_{k}^{s\ast}d_{k}^{s}\exp\left(  i2E_{k}%
^{s}t/\hbar\right)  \right]  .
\end{equation}
For a finite system, $E_{k}^{s}$ belongs to a finite discrete set and the sum
will lead to a time evolution reminiscent of the collapse and revival of Rabi
oscillations. But, in the thermodynamic limit where $V\rightarrow\infty$,
$E_{k}^{s}$ is continuous in $k$ and the discrete sum is transformed into an
integral
\begin{equation}
\dot{n}_{c}\left(  t\right)  \propto\int\frac{k^{2}}{2\pi^{2}}%
\operatorname{Im}\left[  c_{k}^{s\ast}d_{k}^{s}\exp\left(  i2E_{k}^{s}%
t/\hbar\right)  \right]  dk.
\end{equation}
This integral, due to destructive interference, vanishes in the limit
$t\rightarrow\infty$ for any well-behaved $k^{2}c_{k}^{s\ast}d_{k}^{s}$.
Thus, we conclude that a steady-state with $\dot{n}_{c}\approx0$ exists in
the thermodynamic limit.  

Note that in this steady state, $n_{k}(
t)$ displays persistent oscillations in time with the
momentum-dependent frequency $E_{k}^{s}/\hbar$, but this lack of decoherence
is expected within the collisionless Bogoliubov approximation (as is the case
in this study) where real ``collisions" between the Bogoliubov quasi-particles are absent.  At long time, $n_{k}(  t)$ oscillates around its time-averaged value, which is time-independent
and represents the momentum distribution of the nonequilbrium state of the Bose gas in the limit where the quasiparticle interaction is vanishingly small.

How might we characterize this nonequilibrium momentum distribution
[time-averaged $n_{k}(  t)  $], reached after quench?  For an isolated integral
system, Rigol \textit{et al.} \cite{rigol07PhysRevLett.98.050405}, motivated
by a seminal experiment by Kinoshita \textit{et al. }%
\cite{kinoshita06Nature.440.900}, suggested that the steady state be described
by the generalized Gibbs ensemble, which maximizes the many-body von-Neumann
entropy, subject to constraints imposed by all integrals of motion, rather
than described by the standard thermal ensemble, which has only a handful of
integrals of motion, such as total energy, total particle number, etc. This
conjecture is confirmed both numerically \cite{rigol07PhysRevLett.98.050405}
and analytically \cite{cazalilla06PhysRevLett.97.156403} for some correlation
functions in 1-D quantum systems \cite{cazalilla11RevModPhys.83.1405}. 

The present model is different since it consists of two parts: condensate and
quasiparticles. As the two parts interact with each other, the quasiparticle
system is, by itself, not closed, and quantities such as $\hat{\gamma
}_{\mathbf{k}}^{\dag}\hat{\gamma}_{\mathbf{k}}$ are time-dependent. However,
in the limit $t\rightarrow\infty$, the $\hat{\gamma}_{\mathbf{k}}^{\dag}%
\hat{\gamma}_{\mathbf{k}}$ become (approximately) constants of motion, and the
quasiparticle system can be\ regarded as an integral system where the degrees
of freedom are equal in number to the integrals of motion.

We now show analytically that the time average of $n_{k}(  t)  $
at steady state follows exactly the generalized Gibbs ensemble described by
the density operator $\hat{\rho}_{G}=Z_{G}^{-1}\exp[  -\sum_{\mathbf{k}%
}E_{k}^{s}\hat{\gamma}_{\mathbf{k}}^{\dag}\hat{\gamma}_{\mathbf{k}}/(
k_{B}T_{\mathbf{k}})  ]  $, where $Z_{G}=\text{Tr}\exp[
-\sum_{\mathbf{k}}E_{k}\hat{\gamma}_{\mathbf{k}}^{\dag}\hat{\gamma
}_{\mathbf{k}}/(  k_{B}T_{\mathbf{k}})  ]  $ is the partition
function with $k_{B}$ the Boltzmann constant. In the Gibbs ensemble, there
corresponds to each mode $\mathbf{k}$ a temperature $T_{\mathbf{k}}$,
determined by%
\begin{equation}
\left\langle \hat{\gamma}_{\mathbf{k}}^{\dag}\hat{\gamma}_{\mathbf{k}%
}\right\rangle =\left\langle \hat{\gamma}_{\mathbf{k}}^{\dag}\hat{\gamma
}_{\mathbf{k}}\right\rangle _{G}\equiv \text{Tr}\left(  \hat{\gamma}_{\mathbf{k}%
}^{\dag}\hat{\gamma}_{\mathbf{k}}\hat{\rho}_{G}\right)  . \label{G}%
\end{equation}
This may be contrasted with the standard thermal ensemble, which is described
by a unique momentum-independent temperature $T$. A straightforward
application of Eq.\ (\ref{Bogoliubov Transformation}) yields a momentum
distribution, $n_{k}(  t)  \equiv\langle \hat{a}_{\mathbf{k}%
}^{\dag}(  t)  \hat{a}_{\mathbf{k}}(  t)  \rangle
$, in the form of
\begin{align}
n_{k}\left(  t\right)   &  =\left(  u_{k}^{2}+v_{k}^{2}\right)  \left\langle
\hat{\gamma}_{\mathbf{k}}^{\dag}\left(  t\right)  \hat{\gamma}_{\mathbf{k}%
}\left(  t\right)  \right\rangle +v_{k}^{2}\nonumber\\
&\quad  -u_{k}v_{k}\left\langle \hat{\gamma}_{\mathbf{k}}\left(  t\right)
\hat{\gamma}_{-\mathbf{k}}\left(  t\right)  +\hat{\gamma}_{-\mathbf{k}}^{\dag
}\left(  t\right)  \hat{\gamma}_{\mathbf{k}}^{\dag}\left(  t\right)
\right\rangle , \label{nk(t)}%
\end{align}
on one hand, and a momentum distribution in the Gibbs ensemble, $n_{k}(
t)  _{G}\equiv\langle \hat{a}_{\mathbf{k}}^{\dag}(  t)
\hat{a}_{\mathbf{k}}(  t)  \rangle _{G}$, in the form of
\begin{equation}
n_{k}\left(  t\right)  _{G}=\left(  u_{k}^{2}+v_{k}^{2}\right)  \left\langle
\hat{\gamma}_{\mathbf{k}}^{\dag}\left(  t\right)  \hat{\gamma}_{\mathbf{k}%
}\left(  t\right)  \right\rangle _{G}+v_{k}^{2}, \label{nk(t)G}%
\end{equation}
on the other hand. By virtue of the identity in Eq.\ (\ref{G}), the two
distributions are the same apart from the second line of Eq.\ (\ref{nk(t)}).  However, in the limit $t\rightarrow\infty$, using Eq.\ (\ref{c_k and d_k})
and $c_{k}(  t)  \approx c_{k}^{s}\exp[  -iE_{k}^{s}%
t/\hbar]  $ and $d_{k}(  t)  \approx d_{k}^{s}\exp(
iE_{k}^{s}t/\hbar)  $, one can find that this second line equals
$2u_{k}v_{k}\text{Re}(  c_{k}^{s}d_{k}^{s}e^{-i2E_{k}^{s}%
t})  $, which is a periodic function of time and thus averages to zero.  This concludes the proof that the time average of $n_{k}(  t)
$ at steady state is fitted exactly by the generalized Gibbs ensemble distribution.  We stress that this conclusion holds only within the framework of the Bogoliubov theory with well-defined low-energy quasiparticle modes. 

\begin{figure}
[ptb]
\begin{center}
\includegraphics[
width=3.4 in
]%
{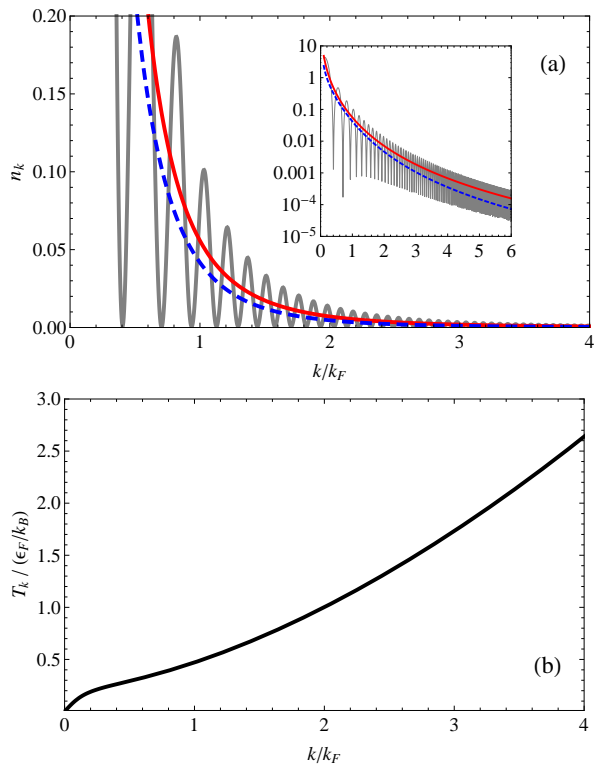}%
\caption{(Color online) (a) The oscillatory and solid curves represent,
respectively, the momentum distribution $n_{k}(  t)  $ and
its time average, evaluated long after the scattering length is quenched from
$a_{i}=0.01n^{-1/3}$ to $a_{f}=0.7n^{-1/3}$. The solid curve also represents
the generalized Gibbs ensemble. The dashed cure is the equilibrium
distribution for a system with permanent scattering length $a_{f}$. The
inset is the same plot, but on a log scale. (b) The temperature $T_{k}$ as a
function of the momentum, where $T_{k}$ is computed using Eq.\ (\ref{beta k}). As in \cite{yin13arXiv:1308.6376}, for all curves (in this article), $a_{f}$ is replaced with Eq.\ (\ref{af}) with $\chi=$ $1$ so that the physics at
unitarity, which is expected to occur in a strongly interacting Bose gas, can be accounted for qualitatively.}%
\label{Fig1}%
\end{center}
\end{figure}

An example of a generalized Gibbs ensemble distribution is displayed by the
solid curve of Fig.\ \ref{Fig1}(a), where the momentum dependent temperature in Fig.\ \ref{Fig1}(b)
fixed by condition (\ref{G}) or explicitly%
\begin{equation}
T_{\mathbf{k}\neq0}=\frac{E_{k}^{s}}{k_{B}\ln\left(  1+\left\vert d_{k}%
^{s}\right\vert ^{-2}\right)  }. \label{beta k}%
\end{equation}
A look at Fig.\ \ref{Fig1}(a) indicates that this nonequilibrium
distribution (solid curve) is different from the equilibrium momentum
distribution (dashed curve). \

\section{Adiabatic Solution and Condition}

The adiabatic theorem, which we discuss in this section, arises
naturally from the time-dependence of the Hamiltonian in Eq.\ (\ref{H2}).
\ We begin by noticing that in the limit where the $\dot{n}_{c}$-induced
interaction is weak in comparison to the quasiparticle energy $E_{k}(
t)  $, or $| U_{k}(  t)  | \ll
E_{k}(  t)  $, Eq.\ (\ref{dot c_k d_k}) supports a particularly
simple solution that Yin and Radzihovsky \cite{yin13arXiv:1308.6376}
discovered
\begin{subequations}
\label{cd}%
\begin{align}
c_{k}\left(  t\right)   &  =c_{k}\left(  0^{+}\right)  \exp\left[  -i\phi
_{k}\left(  t\right)  \right]  ,\\
d_{k}\left(  t\right)   &  =d_{k}\left(  0^{+}\right)  \exp\left[  +i\phi
_{k}\left(  t\right)  \right]  ,
\end{align}
where $\phi_{k}(  t)  =\int_{0}^{t}E_{k}(  t^{\prime})
dt^{\prime}/\hbar$ is the dynamical phase. The momentum distribution within
the validity of this solution becomes%
\end{subequations}
\begin{equation}
n_{k}\left(  t\right)  =n_{k,1}\left(  t\right)  +n_{k,2}\left(  t\right)  ,
\label{nk}%
\end{equation}
where%
\begin{align}
n_{k,1}\left(  t\right)   &  =\epsilon_{k}^{2}\frac{\epsilon_{k}+\left[
\left(  g_{i}+g_{f}\right)  n_{i}+g_{f}n_{c}\left(  t\right)  \right]
}{2E_{k}^{-}E_{k}^{+}E_{k}\left(  t\right)  }\nonumber\\
& \quad +\epsilon_{k}\frac{g_{f}g_{i}n_{i}n_{c}\left(  t\right)  }{E_{k}^{-}%
E_{k}^{+}E_{k}\left(  t\right)  }-\frac{1}{2}, \label{nk1}%
\end{align}
and%
\begin{equation}
n_{k,2}\left(  t\right)  =\frac{\epsilon_{k}\left(  g_{f}-g_{i}\right)
g_{f}n_{i}n_{c}\left(  t\right)  }{E_{k}^{-}E_{k}^{+}E_{k}\left(  t\right)
}\sin^{2}\phi_{k}\left(  t\right)  , \label{nk2}%
\end{equation}
which one can derive by inserting Eq.\ (\ref{cd}) into Eq.\ (\ref{nk(t) 1}).

We comment that one can arrive at the same result by first seeking a solution
in which $n_{c}$ is treated as a constant and then change $n_{c}$ from a
constant to a time-varying parameter $n_{c}(  t)  $ assuming that
the system can instantaneously follow the change in $n_{c}(  t)  $.  For this reason, we compare this simple solution to the adiabatic solution.  This analogy would have been exact (in the usual sense of the adiabatic
solution) had $n_{c}(  t)  $ been a parameter fixed by a source
external to the system.

In the work of \cite{yin13arXiv:1308.6376}, $n_{c}(  t)  $ is
determined self-consistently by combing Eq.\ (\ref{nk}) with the number
equation (\ref{number equation}) or equivalently $n_{c}(  t)  $ can
be integrated from (as we do here),
\begin{equation}
\hbar\frac{dn_{c}\left(  t\right)  }{dt}=\frac{n_{c}\left(  t\right)
\int\frac{dk}{2\pi^{2}}k^{2}\Gamma_{1,k}\sin\left[  2\phi_{k}\left(  t\right)
\right]  }{1+\int\frac{dk}{2\pi^{2}}k^{2}\left\{  \Gamma_{2,k}+\Gamma
_{3,k}\cos\left[  2\phi_{k}\left(  t\right)  \right]  \right\}  },
\label{dnc(t)/dt}%
\end{equation}
which is obtained from Eq.\ (\ref{dnc(t)}) upon neglecting $\dot{n}_{c}$ on its
right-hand side and where $\Gamma_{i,k}\left(  t\right)  $ is given by
\begin{align}
\Gamma_{1,k}\left(  t\right)   &  =-\left(  g_{f}-g_{i}\right)  n_{i}%
g_{f}\frac{\epsilon_{k}}{E_{k}^{-}E_{k}^{+}},\\
\Gamma_{2,k}\left(  t\right)   &  =n_{c}\left(  t\right)  g_{f}^{2}%
\frac{\epsilon_{k}^{2}\left[  \epsilon_{k}+\left(  g_{i}+g_{f}\right)
n_{i}\right]  }{2E_{k}^{-}E_{k}^{+}E_{k}^{3}\left(  t\right)  },\\
\Gamma_{3,k}\left(  t\right)   &  =-\left(  g_{f}-g_{i}\right)  n_{i}%
g_{f}\frac{\epsilon_{k}^{2}\left[  \epsilon_{k}+g_{f}n_{c}\left(  t\right)
\right]  }{2E_{k}^{-}E_{k}^{+}E_{k}^{3}\left(  t\right)  }.
\end{align}
The $\dot{n}_{c}$-induced interaction (\ref{Ut)}) is nothing more than the
adiabatic term, which is expected to arise from the time-dependent Hamiltonian
(\ref{H2}). To illustrate the effect of $\dot{n}_{c}(  t)  $, we
apply the first-order time-dependent perturbation theory in which the term
proportional to $\dot{n}_{c}(  t)$ is treated as a small
parameter. It is then straightforward to show that the ratio between the
$\dot{n}_{c}$-induced interaction, $| i\hbar\epsilon_{k}g_{f}\dot
{n}_{c}(  t)  | /2E_{k}^{2}(  t)  $, and the
quasiparticle energy, $E_{k}(  t)  $,
\begin{equation}
\eta_{k}(  t)  \equiv\frac{\hbar\epsilon_{k}g_{f}\left\vert \dot
{n}_{c}(  t)  \right\vert }{2E_{k}^{3}(  t)  },
\label{eta(k)}%
\end{equation}
represents the first-order correction\ relative to the adiabatic solution
(\ref{cd}) which we now treat as the zeroth order solution in the
perturbation theory.

\begin{figure}
[ptb]
\begin{center}
\includegraphics[
width=3.3in
]%
{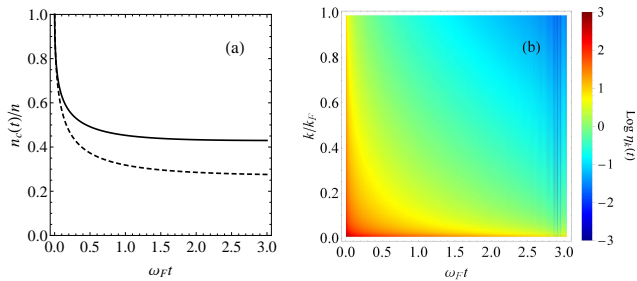}%
\caption{(Color online) Both plots are for the same quench as in Fig.\ \ref{Fig1}. (a) The condensate density $n_{c}\left(  t\right)  $ as a function
of time. The solid curve is from Eqs.\ (\ref{dot c_k d_k}) and (\ref{dot nc}),
where the effect of $\dot{n}_{c}$ is taken into full consideration. This may
be compared with the dashed curve produced from Eq.\ (\ref{dnc(t)/dt}), where
the effect of $\dot{n}_{c}$ is ignored. (b) A contour plot of the
adiabaticity parameter $\eta_{k}\left(  t\right)  $ in Eq.\ (\ref{eta(k)}) as a
function of $k$ and $t$. The regions where $\eta_{k}\left(  t\right)  $ is
larger than one correspond to where the adiabatic approximation fails, which
can be seen to occur most strongly in the small momentum and early time
regions.}
\label{Fig2}%
\end{center}
\end{figure}

$\eta_{k}(  t)  $ in Eq.\ (\ref{eta(k)}) can also be used as a\ nice
figure of merit measuring the adiabaticity of the system when $n_{c}(
t)  $ changes with time. Figure \ref{Fig2}(b) is a contour plot of
$\eta_{k}(  t)  $ using $n_{c}(  t)  $ and $\dot{n}%
_{c}(  t)  $ obtained from Eq.\ (\ref{dnc(t)/dt}) as the
zeroth-order solution.  The formation of a Bose condensate is due to the
spontaneous breaking of a continuous symmetry, which is always accompanied by
a gapless Goldstone mode.  This explains why the adiabatic condition tends
to break down for modes near the Goldstone mode ($k=0$), particularly during
the early stage when $| \dot{n}_{c}(  t)  | $ is
still appreciable.  Thus, Eqs.\ (\ref{nk1}) and (\ref{nk2}) fail to describe
the momentum distribution at small momenta.  Note that $n_{c}(
t)  $ is a collective variable, whose value depends on the histories of
all modes of momenta, be they small or large. As a result, as illustrated in
Fig.\ \ref{Fig2}(a), $n_{c}(  t)  $ integrated from Eq.\
(\ref{dnc(t)/dt}), which uses Eqs.\ (\ref{nk1}) and (\ref{nk2}) as the small
momentum distribution, is noticeably different from $n_{c}(  t)  $
integrated exactly.

We conclude this section by stressing that nonadiabaticity is of particular
importance to the nonequilibrium dynamics we explore in this work. First, we
work with quantum gases consisting of bosons in BEC where, as just discussed,
the instantaneous state defined by the condensate density at time $t$ supports
an excitation spectrum which is always gapless at the Goldstone mode. In
contrast to adiabatic processes involving gapped phases, where the adiabatic
condition may hold when the Hamiltonian changes at a sufficiently slow rate,
here the adiabatic condition breaks down throughout the process at the
Goldstone mode, irrespective of how slowly the condensate density changes with
time. Second, we are interested in the nonequilibrium dynamics of a Bose gas
following a relatively strong interaction quench. The fact that the stronger
the quench, the larger the rate of change in the condensate density $\dot
{n}_{c}\left(  t\right)  $ makes it far more difficult for a strong quench to
maintain the adiabatic condition (\ref{eta(k)}) than a weak quench. This is
the reason that for a strong quench, this nonadiabaticity can have a
relatively large\ effect on measurable quantities such as the condensate
density and Tan's contact, as we illustrate in the rest of the paper.

\section{Condensate Population Dynamics}

\ In this section, we examine more carefully the condensate population
dynamics $n_{c}(  t)  $ and the role that the $\dot{n}$-induced
interaction plays in changing $n_{c}(  t)  $. For simplicity, we
consider the special case, where the quench starts from a BEC in the
non-interacting limit, e.g., $g_{i}=0$ and $n_{i}=n$, and use it as an example
to illustrate the general features of $n_{c}(  t)  $ associated
with any weak-to-strong quench.

First, we seek to quantify $n_{c}(  t)  $ over a time much shorter
than the characteristic relaxation time $\hbar/g_{f}n$, beyond which the
condensate undergoes a significant depletion. We note, in passing, that the
short-time behavior of the condensate fraction and other higher-order
correlation functions in quenched Bose gases have been the focus of several
recent studies
\cite{natu13PhysRevA.87.053607,sykes13arXiv:1309.0828,rancon13PhysRevA.88.031601,rancon14PhysRevA.90.021602,corson14arXiv:1409.0524}%
.$\frac{{}}{{}}$

As a first attempt, we follow the usual time-dependent perturbation theory: we
fix $n_{c}(  t)  $ in Eqs.\ (\ref{dot c_k d_k}) to its initial value
$n$ and propagate Eqs.\ (\ref{dot c_k d_k}) in time. A solution thus derived
will have a momentum distribution $n_{k}(  t)  $ that is nothing
more than Eqs.\ (\ref{nk1}) and (\ref{nk2}) in which $n_{c}(  t)  $
is replaced with $n$. A simple calculation shows that in such a
distribution, $n_{d,1}(  t)  \equiv V^{-1}\sum_{\mathbf{k}}%
n_{k,1}(  t)  $ vanishes so that the condensate depletion becomes
$n_{d}(  t)  =n_{d,2}(  t)  \equiv V^{-1}\sum
_{\mathbf{k}}n_{k,2}ft(  t)  $ or equivalently%

\begin{equation}
n_{d}=g_{f}^{2}n^{2}\int\frac{k^{2}dk}{2\pi^{2}}\frac{\sin^{2}\left[
\sqrt{\epsilon_{k}\left(  \epsilon_{k}+2g_{f}n\right)  }t/\hbar\right]
}{\epsilon_{k}\left(  \epsilon_{k}+2g_{f}n\right)  },
\end{equation}
which can be changed into an integral with respect to $x\equiv k/\sqrt{\hbar
t/2m}$
\begin{equation}
\bar{n}_{d}=\frac{\sqrt{\bar{t}}}{32\pi^{2}\left(  n\xi^{3}\right)  }\int
_{0}^{\infty}dx\frac{\sin^{2}\sqrt{x^{2}\left(  x^{2}+\bar{t}\right)  }}%
{x^{2}+\bar{t}},
\end{equation}
where $\bar{n}_{d}=n_{d}/n$ is a normalized condensate depletion, $\bar
{t}=t/\left(  \hbar/2g_{f}n\right)  $ is the scaled time, and $\xi=\hbar
/\sqrt{4mg_{f}n}$ is the condensate healing length. In the short-time
regime, we find that correct up to order $t^{3/2}$, $\bar{n}_{c}\equiv
n_{c}/n$ changes with time according to%
\begin{equation}
\bar{n}_{c}\left(  t\right)  \approx1+\left[  \bar{n}_{d}\equiv b_{0}\left(
\bar{t}^{1/2}-\frac{\bar{t}^{3/2}}{3}\right)  \right]  ,
\label{nc(t) simple solution}%
\end{equation}
where
\begin{equation}
b_{0}=-\frac{\left(  n\xi^{3}\right)  ^{-1}}{2^{4}\pi^{3/2}}=-4\left(
a_{f}n^{1/3}\right)  ^{3/2}. \label{b0}%
\end{equation}%

\begin{figure}
[ptb]
\begin{center}
\includegraphics[
width=3.5in
]%
{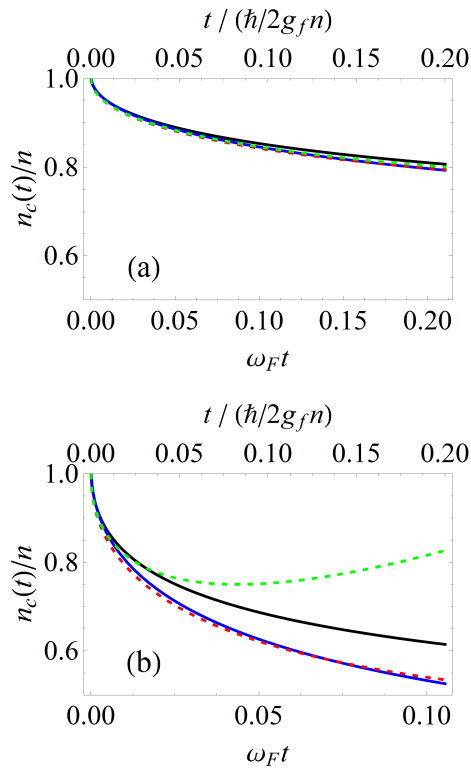}%
\caption{(Color online) A comparison of the short-time $n_{c}(  t)
$ obtained by various methods after a non-interacting BEC is quenched to an
interacting one in which atoms interact with an $s$-wave scattering length (a)
$a_{f}=0.3n^{-1/3}$ and (b) $a_{f}=0.7n^{-1/3}$. The solid black line
represents the exact $n_{c}(  t)  $, the solid blue line the
adiabatic $n_{c}(  t)  $, the dotted green line the simple
perturbation series (\ref{nc(t) simple solution}), and finally the dotted red
line the self-consistent perturbative series (\ref{nc perturbation}). In
(a), the quench is shallow enough that all the lines are virtually on top of
each other.}%
\label{Fig:short-time n_c}%
\end{center}
\end{figure}

Figure \ref{Fig:short-time n_c} compares Eq.\ (\ref{nc(t) simple solution})
(dotted green line) with the adiabatic $n_{c}(  t)  $ (solid blue
line) and the exact $n_{c}(  t)  $ (solid black line). [As a
reminder, we reiterate that the adiabatic $n_{c}(  t)  $ is
integrated from Eq.\ (\ref{dnc(t)/dt}) where the $\dot{n}_{c}(  t)
$-induced interaction is ignored while the exact $n_{c}(  t)  $ is
integrated from Eq.\ (\ref{dot nc}) together with Eqs.\ (\ref{dot c_k d_k})
where the $\dot{n}_{c}(  t)  $-induced interaction is taken into
full consideration.] For a relatively shallow quench in Fig.\
\ref{Fig:short-time n_c}(a) with $a_{f}=0.3n^{1/3}$, up to (at least) $\bar
{t}=0.2$, the simple expansion (\ref{nc(t) simple solution}) agrees nicely
with the two $n_{c}(  t)  $ curves, which themselves agree so well
that they are virtually on top of each other. This is to be contrasted to a
relatively deep quench in Fig.\ \ref{Fig:short-time n_c}(b) with $a_{f}%
=0.7n^{1/3}$, where the simple expansion (dotted blue line), except within a
very short period of time in the order of $\bar{t}=$10$^{-2}$, differs from
others in a quite significant way.

This difference is of no surprise for the following reason. The simple
expansion is integrated from Eqs.\ (\ref{dot c_k d_k}) where $n_{c}(
t)  $ (on their right hand side) is assumed to be \textit{always} fixed
to its initial value $n$. This assumption thus quickly breaks down in a deep
quench where $n_{c}(  t)  $ quickly departs from its initial value
$n$. An improved solution requires $n_{c}(  t)  $ in Eqs.\
(\ref{dot c_k d_k}) to be able to adjust itself in time. Motivated by this
consideration, we develop a self-consistent perturbation theory and use it to
derive from the adiabatic $n_{c}(  t)  $ a perturbative series
\begin{equation}
\bar{n}_{c}\left(  t\right)  \approx1+b_{0}\bar{t}^{1/2}+b_{1}\bar{t}%
+b_{2}\bar{t}^{3/2}, \label{nc perturbation}%
\end{equation}
\ where $b_{0}$ has already been defined in Eq.\ (\ref{b0}), and
\begin{subequations}
\label{b1&2}%
\begin{align}
b_{1}  &  =b_{0}^{2}\left(  1+\frac{\sqrt{\pi}}{8}b_{0}\right)  ,\\
b_{2}  &  =-\frac{b_{0}}{3}+b_{0}^{3}\left(  1+\frac{11\sqrt{\pi}}{32}%
b_{0}+\frac{\pi}{32}b_{0}^{2}\right).
\end{align}
The detailed steps leading to Eq.\ (\ref{nc perturbation}) can be found in the
Appendix. 

For a relatively shallow quench where $b_{0}$ is much less than 1, the terms
linear in $b_{0}$ dominate, and thus, as expected, Eq.\ (\ref{nc perturbation})
reduces to the simple expansion in Eq.\ (\ref{nc(t) simple solution}). In
contrast, for a relatively deep quench where $b_{0}$ is no longer a small
parameter, the higher-order terms of $b_{0}$ become important. It is the
presence of these higher-order terms that allows Eq.\ (\ref{nc perturbation})
[the dotted red line in Fig.\ \ref{Fig:short-time n_c}(b)] to agree, under a
strong quench, with the adiabatic $n_{c}\left(  t\right)  $ (solid blue line)
at a much larger time domain than Eq.\ (\ref{nc(t) simple solution}) (dotted
green line).

We now turn our attention to the exact $n_{c}(  t)  $ which
includes the effect of the $\dot{n}_{c}$-induced interaction. The
interaction in Eq.\ (\ref{Ut)}) is a time-dependent transient function and
is\textit{ imaginary}, introduced to account for the lack of adiabaticity. It serves to damp the initial population buildup in momentum modes,
particularly those with small momenta where the adiabatic condition
(\ref{eta(k)}) is more prone to breaking down. This explains why the exact
$n_{c}(  t)  $ (solid black line) always lags behind and
depletes less deeply than the adiabatic $n_{c}\left(  t\right)  $ (solid blue
line).

This observation leads to a very interesting question. The work of
\cite{yin13arXiv:1308.6376} predicts that for a sufficiently deep quench,
eventually $n_{c}(  t)  $ goes to zero, indicating a phase
transition from a BEC state to a non-BEC state. Since this prediction is
derived from the adiabatic $n_{c}t(  t)  $, will it survive when the
$\dot{n}_{c}$-induced interaction is included?

Before answering this question, we comment on the second component of the
generalized Bogoliubov theory of Yin and Radzihovsky
\cite{yin13arXiv:1308.6376}, in which the scattering length $a_{f}$ is
replaced by the density-dependent scattering amplitude
\end{subequations}
\begin{equation}
a_{f}\rightarrow a_{f}/\sqrt{1+\left(  \chi^{-1}n^{1/3}a_{f}\right)  ^{2}%
}\text{,} \label{af}%
\end{equation}
where $\chi$ is a constant fixed to 1 in their work.  While \textit{ad hoc}, this single function with a single free parameter $\chi$ allows one to capture some qualitative aspects of the physics close to unitarity, where the scattering amplitude is expected to follow the $n^{-1/3}$ universal scaling law.

\begin{figure}
[ptb]
\begin{center}
\includegraphics[
width=3in
]%
{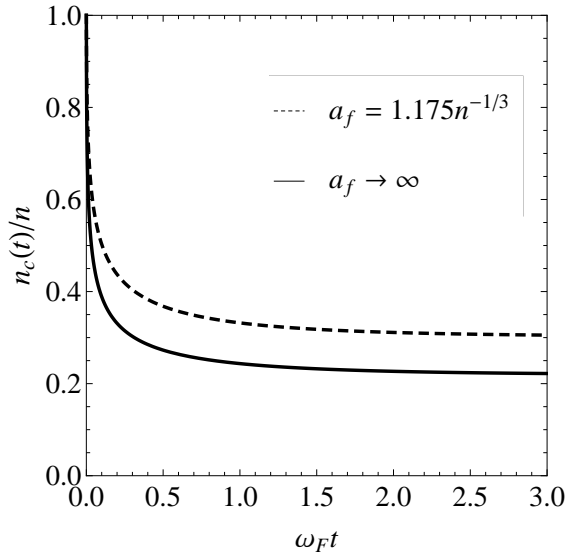}%
\caption{The time evolution of the exact $n_{c}(  t)  $ when
$a_{f}=1.175n^{-1/3}$, given by Eq.\ (\ref{af special}) (dotted line) and the
time evolution of the exact $n_{c}(  t)  $ at unitarity
$a_{f}\rightarrow\infty$ (solid line). In both cases, the quench starts from
a BEC in the non-interacting limit.}%
\label{Fig:nc unitarity}%
\end{center}
\end{figure}

The BEC to non-BEC phase transition is based on the observation that for a
sufficiently deep quench close to resonance, the condensate is so deeply
depleted that the adiabatic $n_{c}(  t)  $ starts to touch zero. In
fact, for a quench that starts from a non-interacting BEC, we find, from atom
number conservation (\ref{number equation}) along with Eq.\ (\ref{af}) with
$\chi=1$, that the adiabatic $n_{c}(  t)  $ approaches a steady
state with vanishing $n_{c}^{s}$ when%
\begin{equation}
a_{f}=\sqrt{\frac{1}{\left(  8/3\sqrt{\pi}\right)  ^{4/3}-1}}n^{-1/3}%
\approx1.175n^{-1/3}, \label{af special}%
\end{equation}
beyond which Eq.\ (\ref{number equation}) at steady state does not support any
positive roots of $n_{c}^{s}$. Instead of depleting to zero, the exact
$n_{c}(  t)  $ at $a_{f}$ $\approx1.175n^{-1/3}$, as shown by the
dotted curve in Fig.\ \ref{Fig:nc unitarity}, approaches a steady state with a
\textit{finite} condensate population, a scenario that continues to hold true
at unitarity where $a_{f}\rightarrow\infty$ as illustrated by the solid curve
in Fig.\ \ref{Fig:nc unitarity}.

The exact $n_{c}(  t)  $ thus does not vanish within the
generalized Bogoliubov theory. This can easily be interpreted as a
consequence of the adiabatic theorem. The adiabatic term or the $\dot{n}%
_{c}$-induced interaction is proportional to $\dot{n}_{c}$, which acts like a
\textquotedblleft velocity\textquotedblright\ dependent friction force.
 This means that the deeper the quench, the faster $n_{c}$ changes, and the
more quickly $n_{c}(  t)  $ slows down. It is precisely 
this mechanism that\ prevents the exact $n_{c}(  t)  $ from being
completely depleted even at unitarity.

\section{Quasiparticle Distribution at High Momenta and Tan's Contact}

The physics at high momenta is closely connected with the physics at short
length scales, which depends on the two-body wave function $\phi(
\mathbf{r})  $, where $\mathbf{r}$ is the relative coordinate [not to be
confused with $\mathbf{r}$ in the many-body Hamiltonian
(\ref{Hamiltonian in position space}), which represents particle position].
 For two particles interacting via a potential having a range $r_{0}$ shorter
than any other length scales, $\phi(  \mathbf{r})  $ obeys the
Bethe-Peierls boundary condition $(  r\phi)  ^{\prime}/(
r\phi)  =-a^{-1}$ or equivalently diverges as $\phi(
\mathbf{r})  $ $\propto r^{-1}-a^{-1}$ in the limit $r\rightarrow0$,
where $a$ is the $s$-wave scattering length. The search for many-body
implications of this two-body physics led Tan
\cite{tan08AnnPhys.323.2952,tan08AnnPhys.323.2971,tan08AnnPhys.323.2987} to
discover, in two-component Fermi gases with zero-range interaction, several
universal relations that link macroscopic quantities, such as the total energy,
to a microscopic quantity that Tan named ``contact," which he defined as the amplitude of the
high-momentum tail of the fermion momentum distribution.

Tan's relations hold irrespective of whether particles are fermions or bosons
and whether they are in equilibrium or nonequilibrium states. We can thus use
our interacting Bose gas as the playground to explore the dynamics of Tan's
contact. We expand the adiabatic distribution in Eqs.\ (\ref{nk1}) and
(\ref{nk2}) at large momenta and find that quasiparticles distribute at large
momenta in the manner of
\begin{equation}
n_{k}^{>}\left(  t\right)  =\frac{C_{k}\left(  t\right)  }{k^{4}}, \label{nk>}%
\end{equation}
where $C_{k}(  t)  $, which is extrapolated from the asymptotic
dependence of $n_{k}(  t)  $ at large momenta, changes with time via
$n_{c}(  t)  $ according to
\begin{align}
C_{k}\left(  t\right)   &  =\left(  \frac{m}{\hbar^{2}}\right)  ^{2}\left[
g_{f}\left(  n_{i}-n_{c}\left(  t\right)  \right)  -g_{i}n_{i}\right]
^{2}\nonumber\\
&\quad  +4\left(  \frac{m}{\hbar^{2}}\right)  ^{2}\left(  g_{f}-g_{i}\right)
g_{f}n_{i}n_{c}\left(  t\right)  \sin^{2}\phi_{k}\left(  t\right)  .
\label{C(t)}%
\end{align}%

\begin{figure}
[ptb]
\begin{center}
\includegraphics[
width=3in
]%
{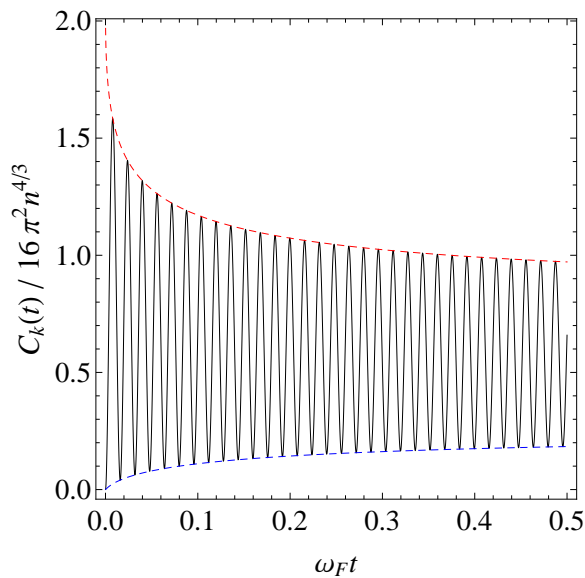}%
\caption{(Color online) The time evolution of $C_{k}(  t)$ in Eq.\ (\ref{C(t)}) at a relatively large momentum after a
non-interacting BEC is quenched to an interacting one. In the particular
example considered here, $k=14k_{F}$ and $a_{f}=1.0n^{-1/3}$. This plot
represents $C_{k}(  t)  $ in Eq.\ (\ref{C(t)}) instead of
$k^{4}n_{k}(  t)  $, where $n_{k}(  t)  $ is given by
Eq.\ (\ref{number equation}), as there exist not much visible difference between the
two calculations within the scale of the plot.}%
\label{Fig;contact dynamics}%
\end{center}
\end{figure}

A salient feature of $n_{k}^{>}(  t)  $ in Eq.\ (\ref{nk>}) is that
$C_{k}(  t)$ in Eq.\ (\ref{C(t)}) encapsulates the dynamics of
$n_{k}^{>}(  t)  $ in all stages. Figure
\ref{Fig;contact dynamics} displays a typical time evolution of $C_{k}(
t)$ at a fixed high momentum. It shows that $C_{k}(  t)
$ develops rapid oscillations that are modulated by a slowly-varying envelop,
with upper bound%
\begin{equation}
C_{\max}\left(  t\right)  =\left(  \frac{m}{\hbar^{2}}\right)  ^{2}\left[
\left(  g_{f}-g_{i}\right)  n_{i}+g_{f}n_{c}\left(  t\right)  \right]  ^{2}
\label{Cmax}%
\end{equation}
and lower bound
\begin{equation}
C_{\min}\left(  t\right)  =\left(  \frac{m}{\hbar^{2}}\right)  ^{2}\left[
g_{f}\left(  n_{i}-n_{c}\left(  t\right)  \right)  -g_{i}n_{i}\right]  ^{2},
\label{Cmin}%
\end{equation}
both of which are independent of the momentum. The frequency of these
oscillations is essentially that of a free particle $\epsilon_{k}/(  2\pi
\hbar)  $. 

As can be seen, immediately following the quench the two envelops scale with
time quite differently, with the upper one changing far more rapidly than the
lower one. The contact dynamics at short times is a reflection of the
short-time behavior of $n_{c}(  t)  $, which was the focus of the
previous section. To gain some insight, we use the short-time\ perturbative
series Eq.\ (\ref{nc perturbation}) and construct a similar expansion for
$C_{\max}(  t)  $ and $C_{\min}(  t)  $. For
simplicity, we only display the expansions through leading order in time,
\begin{equation}
C_{\max}\left(  t\right)  \approx64\pi^{2}a_{f}^{2}n^{2}\left[  1-\sqrt[3]%
{\frac{2^{11}}{3\sqrt{\pi}}}\left(  a_{f}n^{1/3}\right)  ^{2}\sqrt{\omega
_{F}t}\right]  , \label{Cmax(t)}%
\end{equation}
and%
\begin{equation}
C_{\min}\left(  t\right)  \approx64\pi^{2}a_{f}^{2}n^{2}\sqrt[3]{\frac{2^{16}%
}{9\pi}}\left(  a_{f}n^{1/3}\right)  ^{4}\omega_{F}t. \label{Cmin(t)}%
\end{equation}
These results clearly show that $C_{\max}$ and $C_{\min}$ follow different
scaling laws with respect to $t$ (and also $a_{f}n^{1/3}$); in contrast to the
former, which scales with the square root of time, $\sqrt{t}$, the latter
scales linearly with time, $t$.

As expected, close to unitarity, where the interparticle distance, $n^{-1/3}$,
remains the only physically relevant length scale, Eqs.\ (\ref{Cmax(t)}) and
(\ref{Cmin(t)}) approach%

\begin{equation}
C_{\max}\left(  t\right)  \approx64\pi^{2}n^{4/3}\chi^{2}\left(
1-\sqrt[3]{\frac{2^{11}}{3\sqrt{\pi}}}\chi^{2}\sqrt{\omega_{F}t}\right)  ,
\end{equation}
and%
\begin{equation}
C_{\min}\left(  t\right)  \approx64\pi^{2}n^{4/3}\sqrt[3]{\frac{2^{16}}{9\pi}%
}\chi^{6}\omega_{F}t,
\end{equation}
which are functions of the interparticle distance $n^{-1/3}$, independent of
the details\ of the short-range interaction.

Having shed some light on the short-time contact dynamics, we now look at the
long time behavior of the contact. At long times when the system reaches its
steady state, $n_{c}(  t)  \rightarrow n_{c}^{s}$, we can easily
deduce from the time average of Eq.\ (\ref{C(t)}) that%
\begin{equation}
C=\left(  \frac{m}{\hbar^{2}}\right)  ^{2}\left[  \left(  g_{f}-g_{i}\right)
^{2}n_{i}^{2}+g_{f}^{2}n_{c}^{s2}\right]  , \label{C long time}%
\end{equation}
which one may regard as the contact of a nonequilibirum state of a Bose gas in
the limit where the quasiparticle interaction is vanishingly small.  In the absence of a quench, by
setting $g_{i}=g_{f}$ $=g\equiv4\pi\hbar^{2}a/m$ and $n_{c}=n_{i}\equiv n$, we
recover from Eq.\ (\ref{C long time}), the equilibrium contact, $C=(
\frac{m}{\hbar^{2}})  ^{2}g^{2}n^{2}=16\pi^{2}a^{2}n^{2},$ a result
expected from Bogoliubov theory \cite{schakel10arXiv:1007.3452}. 

\begin{figure}
[ptb]
\begin{center}
\includegraphics[
width=3in
]%
{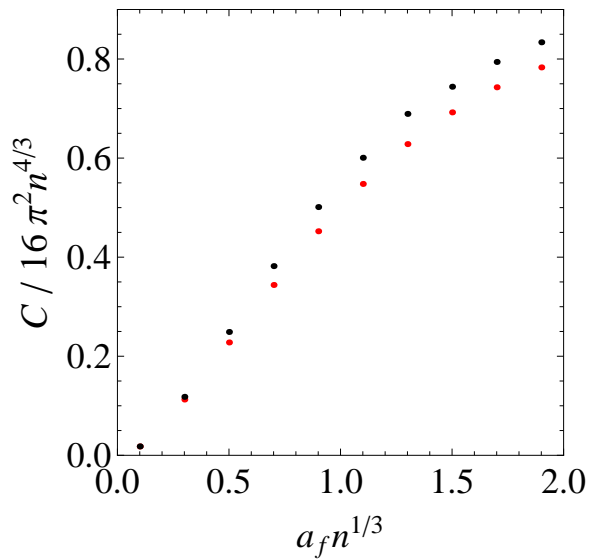}%
\caption{(Color online) The contact $C$ as a function of $a_{f}$ for the
adiabatic $n_{c}^{s}$ [lower (red) points] and for the exact $n_{c}^{s}$
[upper (black) points].}%
\label{Fig:contact}%
\end{center}
\end{figure}

It is important to point out that the contact in Eq.\ (\ref{C long time})
becomes the true contact only when the exact $n_{c}^{s}$ is used. A comparison
between the contact when using the adiabatic $n_{c}^{s}$ and that when using
the exact $n_{c}^{s}$ is given in Fig.\ \ref{Fig:contact}. Again, due to the
slowdown caused by the lack of adiabaticity, the contact based on the exact
$n_{c}^{s}$ (upper points) is higher than the one based on the adiabatic
$n_{c}^{s}$ (lower points), and the difference increases as the level of the
quench increases.%

\section{Conclusion}

In summary, using a self-consistent extension of a Bogoliubov theory
\cite{yin13arXiv:1308.6376} in which the condensate number density $n_{c}$ is
treated as a time-dependent mean field, we constructed a closed set of
equations that highlight the role of $\dot{n}_{c}$, which is to induce an
(imaginary) effective interaction between quasiparticles. We have used this
set of equations to explore the nonequilibrium dynamics of a Bose gas that has
undergone a deep quench to a large scattering length. We have shown
analytically that the system can reach a steady state in which the
time-averaged momentum distribution is described exactly by the generalized
Gibbs ensemble. We studied the adiabatic solution and the conditions upon
which the adiabatic solution holds. We discussed how the $\dot{n}_{c}%
$-induced interaction, which cannot be ignored on the grounds of the adiabatic
approximation for modes near the gapless Goldstone mode, can affect condensate
populations and Tan's contact.

In the course of our study, we constructed a self-consistent perturbation
theory and used it to quantify the adiabatic $n_{c}(  t)  $ at
short times.  We found that while the adiabatic $n_{c}(  t)  $
agrees with the exact $n_{c}(  t)  $ for shallow quenches, the
exact $n_{c}(  t)  $ depletes far less than the adiabatic
$n_{c}(  t)  $ for deep quenches; we found that even when the Bose
gas is quenched to unitarity, the exact $n_{c}(  t)  $ approaches a
steady state with a finite condensate fraction.  We traced this to the
initial suppression of the population buildup in the modes close to the
Goldstone mode, where the adiabatic condition becomes increasingly difficult
to maintain as the level of quench increases. We also used the self-consistent
perturbation series to quantify contact dynamics at short times, finding that
the contact oscillates between two bounds that follow different time scaling
laws.  Finally, we studied the contact of the nonequilibrium state and
found it to be higher in the presence than in the absence of the
$\dot{n}_{c}$-induced interaction.

\section*{Acknowledgments}
B.\ K.\ is grateful to ITAMP and the Harvard-Smithsonian Center for Astrophysics for their hospitality while completing this work.  H.\ Y.\ L is supported in part by the US Army Research Office under Grant No.
W911NF-10-1-0096 and in part by the US National Science Foundation under Grant
No. PHY11-25915.

\appendix

\section{}

In this appendix, we detail the steps taken to extract from Eqs.\
(\ref{number equation}), (\ref{nk1}) and (\ref{nk2}) an analytical series that
accurately approximate $n_{c}(  t)  $ for $t\ll\hbar/g_{f}n$ when
the Bose gas is quenched from a non-interacting BEC state ($g_{i}=0$ and
$n_{i}=n$). While focusing on this limiting case, this appendix also serves
to illustrate the main ingredients and the generality of the self-consistent
perturbation theory which we develop. 

We begin with $n_{d,1}=\frac{1}{V}\sum_{\mathbf{k}}n_{k,1}(  t)  $,
which, when Eq.\ (\ref{nk1}) is used for $n_{k,1}(  t)  $, becomes
\begin{equation}
n_{d,1}=\int_{0}^{\infty}\frac{k^{2}dk}{4\pi^{2}}\left[  \frac{\epsilon
_{k}+g_{f}\left(  n+n_{c}\right)  }{\sqrt{\epsilon_{k}+2g_{f}n}\sqrt
{\epsilon_{k}+2g_{f}n_{c}}}-1\right]  .\label{nd10}%
\end{equation}
Note that in the self-consistent perturbation theory, the condensate density,
$n_{c}(  t)  $, is an unknown function of time to be determined,
but for notational simplicity we will suppress the time dependence unless
confusion is likely to occur. Introducing the relevant healing lengths,
\begin{equation}
\xi=\frac{\hbar}{\sqrt{4mg_{f}n}},
\quad
\xi_{c}=\frac{\hbar}{\sqrt{4mg_{f}n_{c}}%
},\label{r1,r2}%
\end{equation}
we rewrite Eq.\ (\ref{nd10}) into 
\begin{equation}
n_{d,1}=\frac{1}{4\pi^{2}}\int_{0}^{\infty}\left\{  \frac{k^{2}+\frac{1}%
{2}\left(  \xi^{-2}+\xi_{c}^{-2}\right)  }{\sqrt{k^{2}+\xi^{-2}}\sqrt
{k^{2}+\xi_{c}^{-2}}}-1\right\}  k^{2}dk.\label{nd1}%
\end{equation}
During the time regime of interest, the condensate is not significantly
depleted. This allows us to treat%
\begin{equation}
r=\left(  \xi^{-2}-\xi_{c}^{-2}\right)  /\xi^{-2}=1-\bar{n}_{c},\label{r}%
\end{equation}
where $\bar{n}_{c}=n_{c}/n$ is the condensate fraction, as a small
perturbation parameter. Through third order in $r$ (which is sufficient
for a solution valid up to order $t^{3/2}$), Eq.\ (\ref{nd1}) becomes%
\begin{equation}
n_{d,1}\approx\frac{1}{2^{5}\pi^{2}}\int_{0}^{\infty}\left[  \frac{r^{2}%
}{\left(  1+\xi^{2}k^{2}\right)  ^{2}}+\frac{r^{3}}{\left(  1+\xi^{2}%
k^{2}\right)  ^{3}}\right]  k^{2}dk,
\end{equation}
from which we arrive at%
\begin{equation}
n_{d,1}\approx\frac{\xi^{-3}}{2^{7}\pi}\left[  \left(  1-\bar{n}_{c}\right)
^{2}+\frac{\left(  1-\bar{n}_{c}\right)  ^{3}}{4}\right]
,\label{nd1 expansion}%
\end{equation}
where we have replaced $r$ with Eq.\ (\ref{r}).

We now turn our attention to $n_{d,2}=\frac{1}{V}\sum_{\mathbf{k}}%
n_{k,2}(  t)  $, which, when $n_{k,2}(  t)  $ is
substituted with Eq.\ (\ref{nk2}), becomes
\begin{equation}
n_{d,2}=\int\frac{k^{2}dk}{2\pi^{2}}\frac{g_{f}^{2}nn_{c}\sin^{2}\phi
_{k}\left(  t\right)  }{\epsilon_{k}\sqrt{\epsilon_{k}+2g_{f}n}\sqrt
{\epsilon_{k}+2g_{f}n_{c}}}, \label{nd2}%
\end{equation}
where $\phi_{k}(  t)  =\int_{0}^{t}\frac{dt^{\prime}}{\hbar}%
\sqrt{\epsilon_{k}(  \epsilon_{k}+2g_{f}n_{c}(  t^{\prime})
)  }$. The time integration in $\phi_{k}(  t)  $,
unfortunately, cannot be carried out explicitly since we don't know how
$n_{c}(  t)  $ changes with time.  Instead, we approximate
$\phi_{k}(  t)  $ as
\begin{align}
&  \int_{0}^{t}\frac{dt^{\prime}}{\hbar}\sqrt{\epsilon_{k}\left(  \epsilon
_{k}+2g_{f}n_{c}\left(  t^{\prime}\right)  \right)  } \approx\frac{t}{\hbar}\sqrt{\epsilon_{k}\left(  \epsilon_{k}+2g_{f}%
n_{c}\left(  t\right)  \right)  }, \label{phi approximation}%
\end{align}
and will discuss its validity at the end of this appendix. To proceed, we make
the change of variables from $k$ to $x=k\sqrt{\hbar t/(  2m)  }$
and convert Eq.\ (\ref{nd2}) into
\begin{align}
n_{d,2}  &  =\frac{\xi^{-3}\bar{n}_{c}\sqrt{\bar{t}}}{2^{3}\pi^{2}}%
\times  \left(  I=\int_{0}^{\infty}dx\frac{\sin^{2}\sqrt{x^{2}\left(  x^{2}+\bar
{t}\bar{n}_{c}\right)  }}{\sqrt{x^{2}+\bar{t}}\sqrt{x^{2}+\bar{t}\bar{n}_{c}}%
}\right)  , \label{integral I}%
\end{align}
where $\bar{t}=t/(  \hbar/2g_{f}n)  $ is the time relative to the
characteristic relaxation time. An explicit evaluation shows that the
integral in Eq.\ (\ref{integral I}) can be approximated as
\begin{equation}
I\approx\frac{\sqrt{\pi}}{2}+\frac{\sqrt{\pi}}{6}\bar{t}\bar{n}_{c}%
-\frac{\sqrt{\pi}}{3}\bar{t}, \label{I first order}%
\end{equation}
up to the first order in $\bar{t}$, which is all we need in order to yield an
expansion of $n_{c}(  t)  $ up to $t^{3/2}$. Combining Eqs.\
(\ref{nd1 expansion}), (\ref{integral I}) and (\ref{I first order}), we find
$\bar{n}_{c}=1-(  n_{d,1}+n_{d,2})  /n$ or
\begin{align}
\bar{n}_{c}  &  \approx1-\frac{1}{2^{7}\pi\left(  n\xi^{3}\right)  }\left[
\left(  1-\bar{n}_{c}\right)  ^{2}+\frac{\left(  1-\bar{n}_{c}\right)  ^{3}%
}{4}\right] \nonumber\\
&\quad  +\frac{\bar{n}_{c}}{2^{3}\pi^{3/2}\left(  n\xi^{3}\right)  }\left[
\frac{\bar{t}^{1/2}}{2}-\frac{1}{3}\bar{t}^{3/2}+\frac{\bar{n}_{c}}{6}\bar
{t}^{3/2}\right]  . \label{nc}%
\end{align}

If we fix $\bar{n}_{c}$ on the right-hand side of Eq.\ (\ref{nc}) to 1, we
recover the simple perturbative result (\ref{nc(t) simple solution}) in the
main text. In the self-consistent time-dependent perturbation theory we are
developing here, we assume that $\bar{n}_{c}(  t)  $ in Eq.\
(\ref{nc}) is well approximated by the perturbation expansion
\begin{equation}
\bar{n}_{c}\left(  t\right)  \approx1+b_{0}\bar{t}^{1/2}+b_{1}\bar{t}%
+b_{2}\bar{t}^{3/2},\label{nc perturbative ansatz}%
\end{equation}
where $b_{0},$ $b_{1}$, and $b_{2}$ are coefficients to be determined self
consistently. Inserting Eq.\ (\ref{nc perturbative ansatz}) into both sides
of Eq.\ (\ref{nc}) and expanding the right-hand side up to $t^{3/2}$, we have
\begin{align}
&  1+b_{0}\bar{t}^{1/2}+b_{1}\bar{t}+b_{2}\bar{t}^{3/2}\nonumber\\
& \approx1-\frac{\left(  n\xi^{3}\right)  ^{-1}}{2^{7}\pi}\left[  \left(
b_{0}^{2}\bar{t}+2b_{0}b_{1}\bar{t}^{3/2}\right)  -\frac{1}{4}b_{0}^{3}\bar
{t}^{3/2}\right]  \nonumber\\
&\quad  -\frac{\left(  n\xi^{3}\right)  ^{-1}}{2^{3}\pi^{3/2}}\left[  \frac{\left(
\bar{t}^{1/2}+b_{0}\bar{t}+b_{1}\bar{t}^{3/2}\right)  }{2}-\frac{\bar{t}%
^{3/2}}{3}+\frac{\bar{t}^{3/2}}{6}\right]  ,
\end{align}
from which we find, by matching the coefficients,
\begin{align}
b_{0} &  =-\frac{\left(  n\xi^{3}\right)  ^{-1}}{2^{4}\pi^{3/2}},\\
b_{1} &  =\frac{-\left(  n\xi^{3}\right)  ^{-1}}{2^{3}\pi}\left(  \frac
{b_{0}^{2}}{16}+\frac{b_{0}}{2\sqrt{\pi}}\right)  ,\\
b_{2} &  =\frac{-\left(  n\xi^{3}\right)  ^{-1}}{2^{3}\pi}\left[  \frac{b_{0}%
}{16}\left(  2b_{1}-\frac{b_{0}^{2}}{4}\right)  +\frac{b_{1}}{2\sqrt{\pi}%
}-\frac{1}{6\sqrt{\pi}}\right]  .
\end{align}
These results may then be simplified to give Eqs.\ (\ref{b0}) and (\ref{b1&2}). 

The perturbative solution (\ref{nc perturbative ansatz}) made use of the
approximation in Eq.\ (\ref{phi approximation}). This assumption, however,
does not affect the solution (\ref{nc perturbative ansatz}) up to first
order in $t$. To derive the first-order solution, it suffices to replace
$n_{c}(  t^{\prime})  $ and $n_{c}(  t)  $ in Eq.\
(\ref{phi approximation}) with a constant $n$, and under such a circumstance
Eq.\ (\ref{phi approximation}) becomes an equality rather than an approximation and
thus will not have any effect on the validity of the ensuing derivations.
 This is no longer the case for perturbative expansions beyond first
order where $n_{c}(  t)  $ in Eq.\ (\ref{phi approximation}) changes
with time. Nevertheless, Eq.\ (\ref{phi approximation}) remains a good
approximation for modes with kinetic energies higher than $2g_{f}n_{c}(
t)  $. This may partially explain why the solution
(\ref{nc perturbative ansatz}) agrees with the short-time adiabatic
$n_{c}(  t)  $ remarkably well (see Fig.\ \ref{Fig:short-time n_c})
in spite of the approximation we made in Eq.\ (\ref{phi approximation}).



\end{document}